%% file: final_version.tex
\documentclass[journal]{IEEEtran}
\usepackage{multirow}
\usepackage{booktabs}
\usepackage[font=footnotesize]{subcaption}
\usepackage[font=footnotesize]{caption}
\usepackage{amsmath}
\usepackage{amssymb}
\usepackage{ragged2e}
\usepackage{float}
\usepackage{tikz}
\usepackage{cite}
\usepackage[T1]{fontenc}
\usepackage[nolist]{acronym}
\usepackage[linesnumbered,ruled,lined,hanginginout]{algorithm2e}
\usepackage{algpseudocode}

\usepackage{color}
\usetikzlibrary{patterns}

\SetAlgoHangIndent{0em}
\newtheorem{definition}{Definition}
\let\endshdefinition\enddefinition
\def\enddefinition{\strut\hfill$\square$\endshdefinition}

\makeatletter
\g@addto@macro\normalsize{%
  \setlength\abovedisplayskip{3.5pt}
  \setlength\belowdisplayskip{3.5pt}
}

\begin{document}
\IEEEoverridecommandlockouts
	
	\pagenumbering{gobble}
	
	This paper has been accepted for publication on IEEE Wireless Communications Letters.
    
    \copyright 2026 IEEE. Personal use of this material is permitted. Permission from IEEE must be obtained for all other uses, in any current or future media, including reprinting/republishing this material for advertising or promotional purposes, creating new collective works, for resale or redistribution to servers or lists, or reuse of any copyrighted component of this work in other works
	\newpage

\title{Mitigation of UE Antenna Calibration Errors via Differential STBC in Cell-Free Massive MIMO}

\author{Marx M. M. Freitas, \IEEEmembership{Member,~IEEE} and Stefano Buzzi,~\IEEEmembership{Fellow,~IEEE}


\thanks{{This work was supported by the EU under the Italian National Recovery and Resilience Plan (NRRP) of NextGenerationEU, partnership on ``Telecommunications of the Future'' (PE00000001 - program ``RESTART''),  Structural Project NTN, Cascade Call INFINITE, CUP D93C22000910001.}}
\thanks{{The authors are with the Dept. of Electrical and Information Engineering (DIEI), University of Cassino and Southern Lazio, 03043 Cassino, Italy (e-mail:\{marxmiguelmiranda.defreitas; buzzi\}@unicas.it). S. Buzzi is also with the Consorzio Nazionale Interuniversitario per le Telecomunicazioni (CNIT), 43124 Parma, Italy.}}

}
\markboth{}{}

\maketitle

\input{acronym.tex}

\begin{abstract}
This letter investigates the use of differential space–time block coding (DSTBC) to address antenna array calibration impairments at multi-antenna user equipment (UE) in the downlink (DL) of cell-free massive MIMO (CF-mMIMO) systems. 
We show that, by exploiting DSTBC, reliable DL communication can be achieved without explicit UE-side calibration or channel phase knowledge. Simulation results demonstrate that the proposed DSTBC-based transmission effectively mitigates the impact of antenna-dependent phase offsets, restoring near-coherent performance in CF-mMIMO networks.
\end{abstract}

\section{Introduction}

\Ac{UC} \ac{CF-mMIMO} systems are widely regarded as a key enabling technology for next-generation wireless networks \cite{buzziCM2026}. 
Recent \ac{CF-mMIMO} research increasingly focuses on practical hardware impairments, which can severely limit performance if not properly handled. 
Among these investigations, 
non-ideal \ac{RF} chains, whose transmit–receive asymmetries violate the channel reciprocity assumption underpinning \ac{TDD} operation \cite{BookCFemil2021}, are one of the most considered and investigated. Although the wireless channel is reciprocal, hardware distortions create unknown phase and amplitude offsets that conventional precoding techniques cannot remove, degrading coherent \ac{DL} transmission. To address this, intra- and inter-\ac{AP} antenna calibration schemes have been proposed \cite{BeamSyncOverTheAir, OTAEmil2025}, assuming single-antenna \acp{UE}.
These papers address only impairments from the \acp{AP}' \ac{RF} chains and ignore hardware imperfections at the \ac{UE} side.

The performance of \ac{CF-mMIMO} systems further improves when \acp{UE} have multiple antennas, enabling multiple spatial data streams and providing extra array and multiplexing gains \cite{MultiAntennaStefano, kama2025multiantennausers, Shi2024ris}. Hence, antenna calibration at the \ac{UE} side is also relevant. 
While \ac{OTA} calibration is feasible for fixed, network-controlled \acp{AP}, applying it to mobile \acp{UE} is inherently challenging. 
Therefore, this letter studies the unexplored impact of antenna calibration impairments at multi-antenna \acp{UE} in the \ac{DL} of \ac{CF-mMIMO}. Unlike our previous contribution \cite{freitas2025PMCellFree}, which focused on mitigating phase misalignments among distributed \acp{AP}, we show here that \ac{DSTBC} can be also used to overcome channel non-reciprocity caused by unknown, time-varying phase offsets in the \acp{UE}' RF chains, without explicit calibration or phase estimation. Numerical results show that the proposed approach significantly improves \ac{BER} and \ac{SE}, and recovers near-coherent \ac{DL} performance despite UE-side calibration errors.

This letter is organized as follows. Section II describes the system model and its mathematical formulation under channel non-reciprocity. Section III presents the adaptation of \ac{DSTBC} schemes to \ac{CF-mMIMO} systems with multi-antenna \acp{UE} and non-reciprocal channels. Section IV shows numerical results, while, finally,  Section V wraps up the letter.

\section{System Model and Problem Statement}
\label{Sec:SystemModelNew}
We consider a standard \ac{CF-mMIMO} system with $L$ \acp{AP} and $K$ \acp{UE}. Each \ac{AP} has $N_{\mathrm{AP}}$ antennas and each \ac{UE} has $N_{\mathrm{UE}}$ antennas. The \acp{AP} are connected to a \ac{CPU} by error-free fronthaul links, and the system operates in \ac{TDD} mode. The \ac{UL} on-air channel from UE $k$ to AP $l$ is $\mathbf{G}_{k,l} = \beta_{k,l}^{1/2} \mathbf{H}_{k,l}$, where $\beta_{k,l}$ is the large-scale fading coefficient (path-loss and shadowing), and $\mathbf{H}_{k,l} \in \mathbb{C}^{N_{\mathrm{AP}} \times N_{\mathrm{UE}}}$ has \ac{i.i.d} entries distributed as complex Gaussian $\mathcal{N}_{\mathbb{C}}(0,1)$\footnote{$\mathcal{N}_{\mathbb{C}}(\mu,\sigma^{2})$ denotes a complex Gaussian random variable with mean $\mu$ and variance $\sigma^{2}$, and $\mathbb{E}\{\cdot\}$ denotes expectation.}. By channel reciprocity, the \ac{DL} on-air channel from AP $l$ to UE $k$ is $\mathbf{G}_{k,l}^{\mathrm{H}}$. This common assumption allows channel estimation to be done only in the \ac{UL}, with the estimates reused for \ac{DL} transmission within the coherence interval.

\begin{figure}[!]
    \centering
    \includegraphics[width=0.75\linewidth]{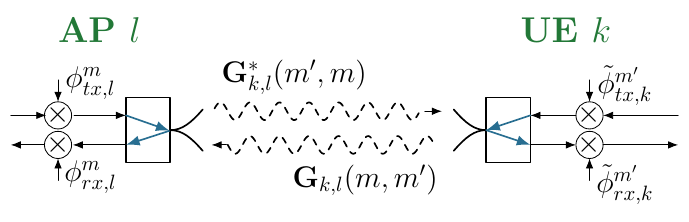}
    \caption{Channel non-reciprocity effects caused by hardware-induced complex-valued offsets between antenna $m$ at \ac{AP} $l$ and antenna $m'$ at \ac{UE} $k$. 
    }
    \label{Fig:HardwareNonReciprocity}
\end{figure}

In practice, however, this assumption is not exact due to hardware non-idealities. As shown in Fig.~\ref{Fig:HardwareNonReciprocity}, the effective channel between \ac{AP} $l$ and \ac{UE} $k$ includes the TX and RX RF branches of each antenna. These hardware-dependent distortions cause amplitude and phase mismatches that break channel reciprocity.
They can be modeled through the diagonal matrices $\mathbf{\Phi}_{rx,l}^{\mathrm{AP}}, \mathbf{\Phi}_{tx,l}^{\mathrm{AP}} \in \mathbb{C}^{N_{\mathrm{AP}} \times N_{\mathrm{AP}}}$ and $\widetilde{\mathbf{\Phi}}_{rx,k}^{\mathrm{UE}}, \widetilde{\mathbf{\Phi}}_{tx,k}^{\mathrm{UE}} \in \mathbb{C}^{N_{\mathrm{UE}} \times N_{\mathrm{UE}}}$, whose $m$-th diagonal elements are the coefficients $\phi_{rx,l}^{m}, \phi_{tx,l}^{m}$ (at AP $l$) and $\widetilde{\phi}_{rx,k}^{m}, \widetilde{\phi}_{tx,k}^{m}$ (at UE $k$), respectively. Consequently, the true \ac{UL} and \ac{DL} channels between the \ac{AP} $l$ and \ac{UE} $k$, within the same coherence block, are: 
\begin{equation}
    \mathbf{\widetilde{G}}_{k,l}^{\mathrm{UL}} = \mathbf{\Phi}_{rx,l}^{\mathrm{AP}} \mathbf{G}_{k,l} \widetilde{\mathbf{\Phi}}_{tx,k}^{\mathrm{UE}}, \quad     \mathbf{\widetilde{G}}_{k,l}^{\mathrm{DL}} = \widetilde{\mathbf{\Phi}}_{rx,k}^{\mathrm{UE}} \mathbf{G}_{k,l}^{\mathrm{H}} \mathbf{\Phi}_{tx,l}^{\mathrm{AP}}.
    \label{Eq:TrueChannels}
\end{equation}
Clearly, we have that $\mathbf{\widetilde{G}}_{k,l}^{\mathrm{UL}} \neq \big[ \mathbf{\widetilde{G}}_{k,l}^{\mathrm{DL}} \big]^{\mathrm{H}}$ as the uncalibrated antenna arrays have destroyed the \ac{UL}/\ac{DL} channel reciprocity. In this letter, these hardware-induced matrices are assumed to be quasi-static over at least two consecutive DSTBC codewords, as RF impairments typically evolve on millisecond timescales, much slower than symbol durations.

To analyze the system impact of uncalibrated antenna arrays, we focus on the \ac{UL} training and DL data transmission phases\footnote{The impact of uncalibrated antenna arrays is less critical in the \ac{UL}, because \ac{UL} channel estimation incorporates their effects and enables their subsequent compensation during the data detection stage.}, with UL channels being estimated via \ac{MMSE} estimation \cite{BeamSyncOverTheAir, freitas2025PMCellFree}.
The generic \ac{UE} $k$ is served by a subset of \acp{AP} called \ac{AP} cluster, denoted as $\mathcal{L}_k \subset \{1,\ldots, L\}$. The AP-UE association policy is encoded in the binary variables $\mathrm{a}_{k,l} \in \{0,1\}$. Since \acp{UE} are equipped with multiple antennas, it is assumed that each \ac{UE} receives $N_s\leq N_{\mathrm{UE}}$ parallel data streams\footnote{For the sake of simplicity, it is assumed that the number of streams $N_s$ is the same for all the \acp{UE}.}. 
We denote by $\mathbf{W}_{k,l}\in \mathbb{C}^{N_{\mathrm{AP}} \times N_{\mathrm{UE}}}$ the precoder used at AP $l$ to transmit data to UE $k$. The latter is built using the \ac{UL} estimated channels. 
Leaving $\mathbf{s}_i^p \in \mathbb{C}^{N_s \times 1}$ be the complex information symbols of the unit-norm intended for \ac{UE} $i$ at a discrete time epoch $p$, the signal received at \ac{UE} $k$ at time $p$ can be expressed as
\begin{equation}
    \mathbf{\widetilde{y}}_k^{p} = \sum\nolimits_{l = 1}^{L} \sum\nolimits_{i = 1}^{K} \mathrm{a}_{i,l} \mathbf{\widetilde{G}}_{k,l}^{\mathrm{DL}} \mathbf{W}_{i,l} \mathbf{M}_i \mathbf{s}_i^{p} + \mathbf{n}_k^{p},
    \label{Eq:ReceivedSignalDL}
\end{equation}
\noindent 
with $\mathbf{n}_k^{p} \in \mathbb{C}^{N_{\mathrm{UE}} \times 1}$ representing the receiver noise. Moreover, $\mathbf{M}_i = \mathbf{I}_{N_s} \otimes \mathbf{1}_{N_\mathrm{UE}/N_s} \in \mathbb{C}^{N_{\mathrm{UE}} \times N_s}$ denotes the channel-independent decoding matrix, where $\otimes$ represents the Kronecker product \cite{MultiAntennaStefano}. In \eqref{Eq:ReceivedSignalDL}, $\mathbf{M}_i$ partitions the UE antennas into disjoint subsets, each processing a distinct data stream (e.g., $N_\mathrm{UE}=6$ and $N_s=3$ yield three subsets). As to the precoding matrix $\mathbf{W}_{i,l}$, one possible choice is:  
\begin{equation}
    \mathbf{W}_{k,l}^{\mathrm{ZISI}} = \sqrt{\rho_{k,l}} \, \mathbf{\widehat{G}}_{k,l}^{\mathrm{UL}} \big(\big(\mathbf{\widehat{G}}_{k,l}^{\mathrm{UL}}\big)^{\mathrm{H}} \mathbf{\widehat{G}}_{k,l}^{\mathrm{UL}}\big)^{-1},
    \label{Eq:PrecodingMatrix}
\end{equation}
where $\mathbf{\widehat{G}}_{k,l}^{\mathrm{UL}} \in \mathbb{C}^{N_{\mathrm{AP}} \times N_{\mathrm{UE}}}$ is the estimated \ac{UL} channel matrix, and $\rho_{k,l}$ is a proper scaling factor that will be specified later.
%
%
As shown in \cite{MultiAntennaStefano}, this beamformer can eliminate inter-stream interference at each UE. We therefore refer to this locally computable precoder as \ac{ZISI}. Given \eqref{Eq:ReceivedSignalDL}, a soft estimate of the data symbols $\mathbf{s}_k^{p}$ is obtained in conventional CF-mMIMO as \cite{MultiAntennaStefano}
\begin{equation}
    \mathbf{\hat{s}}_k^p =  \mathbf{M}_k^{\mathrm{H}} \mathbf{\widetilde{y}}_k^{p},
    \label{Eq:DecodingDataBuzzi}
\end{equation}
where $\mathbf{M}_k$ combines signals from the corresponding antenna subsets to process each data stream separately. Substituting \eqref{Eq:ReceivedSignalDL} and \eqref{Eq:PrecodingMatrix} into \eqref{Eq:DecodingDataBuzzi}, and assuming that the channel is perfectly estimated, after some manipulations, $\mathbf{\hat{s}}_k^p$ becomes
\begin{equation}
\begin{array}{llll}
     \!  \displaystyle
    \sum_{l \in \mathcal{L}_k} \sqrt{\rho_{k,l}} \mathbf{M}_k^{\mathrm{H}} \mathbf{\widetilde{G}}_{k,l}^{\mathrm{DL}}  \mathbf{\widetilde{G}}_{k,l}^{\mathrm{UL}} \big[\big(\mathbf{\widetilde{G}}_{k,l}^{\mathrm{UL}}\big)^{\mathrm{H}} \mathbf{\widetilde{G}}_{k,l}^{\mathrm{UL}}\big]^{-1}\! \mathbf{M}_k \mathbf{s}_k^{p} \! + \mathbf{i}_k^p .
\label{eq:received_calibrated}    
\end{array}\end{equation}
In \eqref{eq:received_calibrated}, we have isolated the useful term, whereas all interference and additive noise affecting UE $k$ have been grouped into the term $\mathbf{i}_k^p$. 
One can note that when there are no antenna calibration errors (i.e., when $\mathbf{\widetilde{G}}_{k,l}^{\mathrm{DL}} = [\mathbf{\widetilde{G}}_{k,l}^{\mathrm{UL}}]^{\mathrm{H}}$), the useful term in \eqref{eq:received_calibrated} simplifies to $(N_\mathrm{UE}/N_s)\sum_l a_{k,l} \sqrt{\rho_{k,l}} \mathbf{s}_k^{p}$, which is thus proportional to the data vector to be estimated. In the more realistic scenario where calibration errors occur, \eqref{eq:received_calibrated} instead indicates that the data vector $\mathbf{s}_k^{p}$ is premultiplied by a non-diagonal mixing matrix, which significantly degrades the system performance.
In particular, assuming that the antenna arrays at the \acp{AP} are perfectly calibrated \footnote{In practice, AP arrays are not perfectly calibrated. In this letter, we assume perfectly calibrated APs to isolate the effects of UE-side calibration errors. AP calibration procedures are available in, e.g.,  \cite{BeamSyncOverTheAir, OTAEmil2025}.}, i.e. that $\mathbf{\Phi}_{rx,l}^{\mathrm{AP}}$ and $\mathbf{\Phi}_{tx,l}^{\mathrm{AP}}$ are identity matrices, for any $l$, the true \ac{UL} and \ac{DL} channels reduce to $\mathbf{\widetilde{G}}_{k,l}^{\mathrm{UL}} = \mathbf{G}_{k,l} \widetilde{\mathbf{\Phi}}_{tx,k}^{\mathrm{UE}}$ and $\mathbf{\widetilde{G}}_{k,l}^{\mathrm{DL}} = \widetilde{\mathbf{\Phi}}_{rx,k}^{\mathrm{UE}} \mathbf{G}_{k,l}^{\mathrm{H}}$, respectively. Substituting these expressions into \eqref{eq:received_calibrated} and applying the identity $\big[
(\widetilde{\mathbf{\Phi}}_{tx,k}^{\mathrm{UE}})^{\mathrm{H}}
\mathbf{A}_{k,l}
\widetilde{\mathbf{\Phi}}_{tx,k}^{\mathrm{UE}}
\big]^{-1}
=
(\widetilde{\mathbf{\Phi}}_{tx,k}^{\mathrm{UE}})^{-1}
\mathbf{A}_{k,l}^{-1}
(\widetilde{\mathbf{\Phi}}_{tx,k}^{\mathrm{UE}})^{-\mathrm{H}}$, with $\mathbf{A}_{k,l} = \mathbf{G}_{k,l}^{\mathrm{H}} \mathbf{G}_{k,l}$, the useful term in \eqref{eq:received_calibrated} becomes 
\begin{equation}
    \displaystyle
    \sum\nolimits_{l = 1}^{L} \!\mathrm{a}_{k,l} \sqrt{\rho_{k,l}} \mathbf{M}_k^{\mathrm{H}} 
    \widetilde{\mathbf{\Phi}}_{rx,k}^{\mathrm{UE}} \big[\widetilde{\mathbf{\Phi}}_{tx,k}^{\mathrm{UE}} \big]^\mathrm{-H}   
    \mathbf{M}_k \mathbf{s}_k^{p} \;,
    \label{eq:usefuldata_uncalibrated}
\end{equation}
where $(\cdot)^{-\mathrm{H}}$ is the Hermitian inverse. Eq.\,\eqref{eq:usefuldata_uncalibrated} reveals that, even under perfect UL channel estimation and perfectly calibrated antenna arrays at the APs, the UEs' RF chains still introduce distortions in the received signal. Moreover, for each data stream, the contributions from the $(N_\mathrm{UE}/N_s)$ antennas assigned to that stream are combined in a non-coherent manner due to the unknown distortions affecting the UE RF chains. In the following, we show how DSTBC can be incorporated into the proposed system to mitigate these effects on data detection without requiring knowledge of $\widetilde{\mathbf{\Phi}}_{rx,k}^{\mathrm{UE}}$ and $\widetilde{\mathbf{\Phi}}_{tx,k}^{\mathrm{UE}}$.

\section{DSTBC-based DL Data Transmission}
\label{sec:DSTBC_design}

DSTBC extends classical STBC by transmitting multiple copies of a user’s data symbols across antennas and time, encoding them into structured space–time codewords. The receiver recovers the data from phase differences between successive codewords, thus obtaining transmit diversity without requiring instantaneous channel state information. In \cite{freitas2025PMCellFree}, the authors showed how \ac{DSTBC} can address \ac{AP} phase misalignments in joint coherent \ac{DL} data transmission for single-antenna \acp{UE}. In this letter, we instead use \ac{DSTBC} to handle \ac{UE}-side antenna calibration errors. We propose a transmission scheme that removes the need to calibrate \ac{UE} antennas and mitigates distortions from the \acp{UE}’ \ac{RF} chains. We also design the differential encoding, transmission, and data detection to properly combat \ac{UE}-side antenna calibration errors.

\subsection{Differential Encoding and Transmission}
Consider an arbitrary UE $k$ served by a subset of APs denoted by $\mathcal{L}_k$, with $L_k = |\mathcal{L}_k|$ being the number of APs serving this UE. Let $P$ represent the number of symbol periods spanned by each space-time codeword. Recalling that $N_s$ parallel data streams are intended for UE $k$, denote by $\{ \mathcal{S}_{k,j}\}_{j=1}^{N_s}$ the collection of symbol sequences for these streams. 
Specifically, the $j$th data stream consists of the sequence $\mathcal{S}_{k,j} = \{ s^{1}_{k,j},\, s^{2}_{k,j},\, \dots,\, s^{N_{\text{sym}}}_{k,j} \}$, 
where $N_{\text{sym}}$ is the total number of complex symbols for stream $j$, each drawn from a \textit{unitary} constellation. 
The central processing unit (CPU) splits each stream $\mathcal{S}_{k,j}$ into smaller segments of $n_s$ symbols each. Denoting by $\tau_d$ the length in discrete samples of the channel coherence time devoted to data transmission, we let $G = \left\lfloor \tau_d / P \right\rfloor-1$ denote the number of such segments within the $\tau_d$ data-symbol duration, so that $N_{\text{sym}} = (G-1)n_s$.
The $t$-th segment of stream $j$ is denoted by 
\begin{equation}
    \mathcal{S}^{t}_{k,j} = \{\, s^{(t-1)n_s + 1}_{k,j},\ \dots,\ s^{t\,n_s}_{k,j} \,\}, 
    \quad |\mathcal{S}^{t}_{k,j}| = n_s,
    \label{eq:stream_segment}
\end{equation}
for $t=1,2,\dots,G-1$. Each such symbol segment is then mapped to a space--time code matrix $\mathbf{X}^t_{k,j} \in \mathbb{C}^{\,L_k \times P}$. 
For simplicity and without loss of generality, we focus on \textit{square} STBC codewords, i.e., we assume that $L_k = P$. This assumption holds for many standard orthogonal designs.

Once the set of code matrices $\{\mathbf{X}^t_{k,j}\}$ is defined for each data stream $j$, the CPU performs \textit{differential encoding} across successive codewords. In particular, the CPU generates a sequence of transmit matrices $\{\mathbf{C}^t_{k,j}\}$ by the recursion 
\begin{equation}
    \mathbf{C}^t_{k,j} \;=\; \mathbf{C}^{\,t-1}_{k,j}\; \mathbf{X}^t_{k,j}\,,
    \qquad \mathbf{C}^0_{k,j} = \mathbf{I}_{L_k}\,,
    \label{eq:differential_encoding}
\end{equation}
for each stream $j=1,\dots,N_s$. The matrix $\mathbf{C}^t_{k,j}\in\mathbb{C}^{\,L_k\times L_k}$ represents the encoded information to be transmitted to UE $k$ on stream $j$ during the $t$-th codeword interval. 
To this aim, the proposed scheme assumes that each row of $\mathbf{C}^t_{k,j}$ is assigned to one of the APs serving UE $k$. 
Since the APs have their own indexing, we define a mapping function $m(l,k)$ that maps each actual AP index $l \in \mathcal{L}_k$ to a row index in $\{1,\dots,L_k\}$. Before transmission, the CPU uses this mapping to distribute the rows of $\mathbf{C}^t_{k,j}$ to the appropriate APs. Specifically, the row of $\mathbf{C}^t_{k,j}$ assigned to AP $l$ (where $l \in \mathcal{L}_k$) is given by\footnote{Eq.\,\eqref{eq:row_mapping} remains valid if $L_k < P$. For instance, in a configuration with $L_k = 3$ and $P = 4$, the \ac{CPU} does not assign the last row of $\mathbf{C}^t_{k,j}$ to any \ac{AP}, as Eq.\,\eqref{eq:row_mapping} maps the rows of $\mathbf{C}^t_{k,j}$ only to the \acp{AP} serving the UE.}
\begin{equation}
    \mathbf{c}^t_{k,l,j} \;=\; \big[\,\mathbf{C}^{\,t-1}_{k,j}\big]_{\,m(l,k),:\;} \; \mathbf{X}^t_{k,j}\,,
    \label{eq:row_mapping}
\end{equation}
which is a $1\times L_k$ row-vector. In other words, AP $l$ (the $m(l,k)$-th serving AP for UE $k$) will transmit the row $\mathbf{c}^t_{k,l,j}$ of the new code matrix $\mathbf{C}^t_{k,j}$ for each stream $j$. The collection of signals to be sent by AP $l$ to UE $k$ (across all streams) can be gathered into an $N_s \times L_k$ matrix 
\begin{equation}
    \mathbf{B}^t_{k,l} \;=\; \begin{bmatrix} 
        (\mathbf{c}^t_{k,l,1})^{T}  \ldots  (\mathbf{c}^t_{k,l,N_s})^{T} 
    \end{bmatrix}^{T}\!,
    \label{eq:B_matrix}
\end{equation}
which stacks the transmit row-vectors for streams $1, \ldots, N_s$. 
The $l$-th \ac{AP} then precodes and transmits the signal blocks $\mathbf{B}^t_{k,l}$ to the UE $k$ using the beamformer $\mathbf{W}_{k,l} \in \mathbb{C}^{\,N_{\text{AP}}\times N_{\text{UE}}}$ introduced in the previous section. The signal received by UE $k$ over the $t$-th codeword interval is thus expressed as
\begin{equation}
    \mathbf{Y}^t_k \;=\; \sum\nolimits_{l \in \mathcal{L}_k} \widetilde{\mathbf{G}}^{\text{DL}}_{k,l}\, \mathbf{W}_{k,l}\, \mathbf{M}_k\, \mathbf{B}^t_{k,l} \;+\; \mathbf{N}^t_k\,,
    \label{eq:received_block}
\end{equation}
where  $\mathbf{N}^t_k \in \mathbb{C}^{N_{\mathrm{UE}} \times L_k}$ in \eqref{eq:received_block} is the combined interference-plus-noise matrix at UE $k$ during the $t$-th codeword interval. 
The key challenge is that UE $k$ does not know the effective channel $\widetilde{\mathbf{G}}^{\text{DL}}_{k,l} \mathbf{W}_{k,l} \mathbf{M}_k$ in \eqref{eq:received_block}; this motivates a differential detection approach, as described next.

\subsection{Differential Detection at the Receiver}
At UE $k$, the goal is to detect the transmitted code matrices $\mathbf{X}^t_{k,j}$ without knowing the instantaneous channel phases. To do so, 
we assume that the unknown offset matrices $\widetilde{\mathbf{\Phi}}_{tx,k}^{\mathrm{UE}}$ and $\widetilde{\mathbf{\Phi}}_{rx,k}^{\mathrm{UE}}$ remain approximately \emph{constant}\footnote{This assumption is practical as phase offsets from calibration impairments change more slowly than baseband symbols, thus remaining quasi-static over two differential blocks. For example, 5G NR symbols last tens of microseconds, whereas hardware offsets vary over milliseconds or longer.}
over at least two consecutive codeword intervals. Under this assumption, the effective channel matrices for intervals $t$ and $t-1$ are basically the same. Thus the channel effects cancel out when processing $\mathbf{Y}^t_k$ and $\mathbf{Y}^{\,t-1}_k$ differentially, as briefly illustrated in Appendix B.
Because multiple data streams are transmitted (one per UE antenna group), the $N_{\text{UE}}$ antennas at UE $k$ observe a superposition of all streams. From \eqref{eq:usefuldata_uncalibrated}, each stream is received over $(N_\mathrm{UE}/N_s)$ antennas. Thus, the $N_{\text{UE}}$ antennas can be conceptually divided into $N_s$ groups of $N_b = N_{\text{UE}}/N_s$ antennas, each mainly corresponding to one stream. To separate the streams, UE $k$ partitions the received signal matrix $\mathbf{Y}^t_k$ by grouping its rows per stream. In particular, the portion of $\mathbf{Y}^t_k$ carrying stream $j$ is obtained by extracting the rows associated with the $j$-th antenna group. We denote this sub-matrix as 
\begin{equation}
    \mathbf{\widetilde{Y}}^t_{k,j} \;\triangleq\; \big[\,\mathbf{Y}^t_k\,\big]_{(j-1)N_b + 1 :\; jN_b,\ :}\,,
    \label{eq:stream_extraction}
\end{equation}
for $j=1,\dots,N_s$, where the notation $[\,\cdot\,]_{a:b,\,:}$ indicates we take rows $a$ through $b$ of $\mathbf{Y}^t_k$ (and all columns). An analogous definition gives $\mathbf{\widetilde{Y}}^{\,t-1}_{k,j}$ from $\mathbf{Y}^{\,t-1}_k$. After this stream-wise separation, each $\mathbf{\widetilde{Y}}^t_{k,j}$ is an $N_b \times L_k$ matrix containing the received superimposed codeword for stream $j$ (from all $L_k$ serving APs) over the $P=L_k$ symbol periods.

Finally, assuming that \ac{PSK} modulation is employed, the detection of $\mathbf{X}^{t}_{k,j}$ can be performed using the following \ac{ML} criterion \cite{Larsson_Stoica_2003}
\begin{equation}
    \centering
    \hat{\mathbf{X}}^{t}_{k,j} \!=\! \mathop{\mathrm{arg\,max}}_{\mathbf{X} \in \mathcal{X}_{N}} \, \mathrm{Re} \big \{ \mathrm{tr} \big \{  \mathbf{X} \, \big ( \mathbf{\widetilde{Y}}^t_{k,j} \big)^{\mathrm{H}} \mathbf{\widetilde{Y}}^{t-1}_{k,j} \big \} \big \},
    \label{eq:DifferentialDecodingTraceRealMin}
\end{equation}
where $\mathcal{X}_{N}$ is the set of all possible code matrices for a unitary constellation $\mathcal{S}$. 
In \eqref{eq:DifferentialDecodingTraceRealMin}, $\mathrm{Re} \{\cdot\}$ denotes the real part, and $\mathrm{tr}\{\cdot\}$ is the matrix trace. This criterion compares the correlation between the received signal blocks $\mathbf{\widetilde{Y}}^t_{k,j}$ and $\mathbf{\widetilde{Y}}^{t-1}_{k,j}$ with each candidate code matrix $\mathbf{X}$ in the codebook, and selects the one that best matches. Although the product $(\mathbf{\widetilde{Y}}^t_{k,j})^{\mathrm{H}} \mathbf{\widetilde{Y}}^{t-1}_{k,j}$ may introduce multi-user interference cross-terms, their impact is reduced by the adopted precoding techniques\footnote{{In the multi-user case, the ML detector in \eqref{eq:DifferentialDecodingTraceRealMin} is applied as in the single-user setting reported in \cite{Larsson_Stoica_2003}, where the multi-user interference is treated as part of the effective noise $\mathbf{N}^t_k$ in \eqref{eq:received_block}.}}. The entire outlined procedure is summarized in Algorithm\,\ref{algo1}.

\begin{algorithm}[htb!]

\DontPrintSemicolon 

\KwIn{$N_s$, $G$, $t = 1,\ldots,G-1$, $k = 1,\ldots,K$;}

\tcp*[h]\CommentSty{Processing at the CPU}

The CPU receives $\{ \mathcal{S}_{k,j}\}_{j=1}^{N_s}$; \tcp*[h]\CommentSty{Parallel data streams}

\For{$j = 1 \; \mathbf{to} \; N_s$}{
Split $\mathcal{S}_{k,j}$ into segments $\mathcal{S}^{t}_{k,j}$ in \eqref{eq:stream_segment}; 

Map $\mathcal{S}^{t}_{k,j}$ onto a space-time code matrix $\mathbf{X}^t_{k,j}$;

Differential encoding: $\mathbf{C}^t_{k,j} = \mathbf{C}^{\,t-1}_{k,j}\; \mathbf{X}^t_{k,j}$ in \eqref{eq:differential_encoding};

Assign the rows of $\mathbf{C}^t_{k,j}$ to the serving APs via \eqref{eq:row_mapping};
}

Build $\mathbf{B}^t_{k,l}$ in \eqref{eq:B_matrix} and forward it to the serving APs;

The serving APs transmit $\mathbf{B}^t_{k,l}$ to UE $k$;

\tcp*[h]\CommentSty{Data detection at the UE}

UE $k$ receives the aggregate DL signal $ \mathbf{Y}^t_k$ in \eqref{eq:received_block};
    
\For{$j = 1 \; \mathbf{to} \; N_s$}{

Extract $\mathbf{\widetilde{Y}}^t_{k,j}$ and $\mathbf{\widetilde{Y}}^{t-1}_{k,j}$ from $\mathbf{Y}^t_k$ and $\mathbf{Y}^{t-1}_k$ in \eqref{eq:stream_extraction};

Detect $\mathbf{X}^t_{k,j}$ via \eqref{eq:DifferentialDecodingTraceRealMin}, i.e., $\hat{\mathbf{X}}^{t}_{k,j}$;
}

\KwOut{$\hat{\mathbf{X}}^{t}_{k,j}$}
\caption{DSTBC-based DL Data Transmission}
    \label{algo1}

\end{algorithm}

\subsection{Complexity Analysis and System Overhead}

The proposed scheme adds complexity from differential encoding at the CPU and data detection at the UE. At the CPU, \eqref{eq:differential_encoding} requires one $L_k \times L_k$ matrix multiplication per stream, i.e., $L_k^3$ complex multiplications. Since $L_k$ is typically small (e.g., $L_k \leq 4$), this cost is negligible compared to the precoding overhead in coherent transmission.

At the UE, the main operations are: (i) computing $(\mathbf{\widetilde{Y}}^t_{k,j})^{\mathrm{H}} \mathbf{\widetilde{Y}}^{t-1}_{k,j}$, requiring $N_b L_k^2$ complex multiplications per stream; and (ii) the codebook search. Although the codebook has size $M^{n_s}$, the orthogonality of the code matrices allows symbol-wise decoding, reducing complexity from $\mathcal{O}(M^{N_s})$ to $\mathcal{O}(N_s M)$. Thus, as in coherent detection, the overall complexity stays linear in the constellation size \cite{Larsson_Stoica_2003}.

\section{Numerical Results}
Unless stated otherwise, we consider a \ac{CF-mMIMO} network comprising $K = 20$ \acp{UE} and $L = 40$ \acp{AP}. The \acp{UE} are equipped with $N_{\mathrm{UE}} = 2$ antennas, while each \ac{AP} has $N_{\mathrm{AP}} = 8$ antennas. Each \ac{UE} receives $N_s = 2$ simultaneous data streams. The \acp{UE} are uniformly distributed within a  $0.5 \times 0.5$ km$^2$ area, while the distribution of \acp{AP} follow a \ac{HCPP}\footnote{This approach enforces a minimum distance of $d_{\text{min}} = \sqrt{A/L}$ between any two \acp{AP}, where $A$ denotes the coverage area. 
}.
The entries of the matrices $\widetilde{\mathbf{\Phi}}_{tx,k}^{\mathrm{UE}}$ and $\widetilde{\mathbf{\Phi}}_{rx,k}^{\mathrm{UE}}$ are assumed to have unit modulus and uniformly distributed phase, for any $k$. Besides the \ac{ZISI} precoder, also the \ac{P-MMSE} precoder, described in Appendix A, is used for the numerical results.
We consider a channel coherence block of discrete length $\tau_{c} = 200$; of these $\tau_{p} = 16$ samples are used for \ac{UL} training, and $\tau_{d} = 184$ for \ac{DL} transmission. The total transmit powers per \ac{UE} and per \ac{AP} are set to $100\,\mathrm{mW}$ and $200\,\mathrm{mW}$, respectively.

To evaluate the \ac{BER} and \ac{SE}, Monte Carlo simulations are performed to account for different channel realizations and \ac{AP}/\ac{UE} locations, referred to as setups. 
{The approximated \ac{SE} for \ac{UE}~$k$ is computed as a BER-based throughput as}
\begin{equation}
    \centering
    \mathrm{SE}_k = P_f \log_2 \left( M_o \right) \left(1-\mathbb{E}\left\{ \mathrm{BER}_k \right \} \right)
\end{equation}
\noindent where $M_o = 8$ is the modulation order, and $\mathbb{E}\left\{ \mathrm{BER}_k \right \}$ denotes the average \ac{BER} computed over all channel realizations. $P_f$ is the usual pre-log factor; specifically, $P_f = \tau_d / \tau_c$ for conventional \ac{CF-mMIMO}, and $P_f = (G - 1)n_s / \tau_c$ for \ac{CF-mMIMO} employing \ac{DSTBC}, with $G = \lfloor \tau_d / L_k \rfloor$. {Let $\mathcal{K}_l \subset \{1,\ldots, K\}$ denote the subset of \acp{UE} served by \ac{AP} $l$.} In distributed processing, the power allocated to \ac{UE} $k$ regarding \ac{AP} $l$ is given by $\rho_{k,l} = \rho_{k,l}^{norm} \big(\sqrt{\beta_{k,l}}/ \sum_{k^{\prime} \in \mathcal{K}_l} \sqrt{\beta_{k^{\prime},l}}\big)$, where $\rho_{k,l}^{norm}=\rho_{max}/\big\| \mathbf{W}_{k,l}^{\mathrm{ZISI}} \big\|^{2}_{F}$ and $k \!\neq \! k^{\prime}$. 
In the centralized one, the power coefficient $\rho_k$ is computed as \cite{Chen2023, kama2025multiantennausers}
\begin{equation}
\rho_k=\frac{\rho_{\max}}{\rho_k^{norm}} \frac{\Big(\sum_{l \in \mathcal{L}_k} \beta_{k, l}^{\varsigma}\Big)^\kappa}{\max _{\ell \in \mathcal{L}_k} \sum_{i \in \mathcal{K}_\ell}\Big(\sum_{l \in \mathcal{L}_i} \beta_{i, l}^{\varsigma}\Big)^\kappa  } ,
\end{equation}
where $\rho_k^{norm} = \sum_{l=1}^{L}\mathbb{E}\big\{\big\| \mathbf{W}_{k,l}^{\mathrm{P-MMSE}} \big\|^{2}_{F}\}$, and the operator $\| \cdot \|_{F}$ represents the Frobenius norm. Moreover, $\varsigma$ and $\kappa \in [0, 1]$ are design parameters, defined as $\varsigma = 0.2$ and $\kappa = 0.5$ \cite{BookCFemil2021, Chen2023}. The 3GPP \ac{UMi} path loss model is used to describe the propagation channel. 
The shadow fading standard deviation is $4\,\mathrm{dB}$; AP and UE antenna heights are $11.65\,\mathrm{m}$ and $1.65\,\mathrm{m}$, respectively; the receiver noise figure is $8\,\mathrm{dB}$. The system uses a $3.5\,\mathrm{GHz}$ carrier with $20\,\mathrm{MHz}$ bandwidth.
For \ac{AP} clustering, we associate each \ac{UE} with the $L_k$ \acp{AP} with the strongest channel gains \cite{RestrictedProcessingMarx}, while 
pilot assignment follows \cite{BookCFemil2021}.

\begin{figure}[htb!]
    \centering
    \begin{subfigure}[b]{0.238\textwidth}
        \centering
        \hspace{0.4cm} \includegraphics[width=3.25cm]{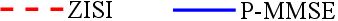}
        \vspace{0.03cm}
        \centering
        \includegraphics[width=\linewidth]{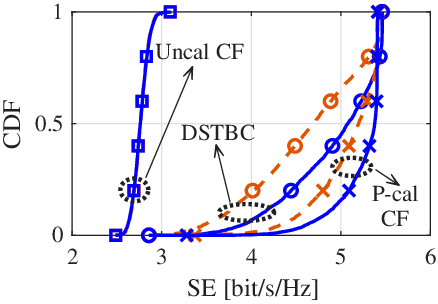}
        \caption{SE, $L_k = 2$ \acp{AP}}
        \label{Eq:CDF_a}
    \end{subfigure}
    \hspace{-0.2cm}
    \begin{subfigure}[b]{0.238\textwidth}
        \centering
        \hspace{0.4cm} \includegraphics[width=3.25cm]{./Figures/Legends.eps}
        \vspace{0.03cm}
        \centering
        \includegraphics[width=\linewidth]{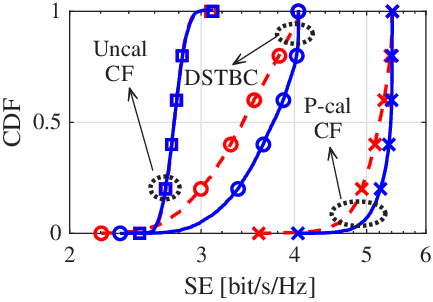}
        \centering
        \caption{SE, $L_k = 4$ \acp{AP}}
        \label{Eq:CDF_b}
    \end{subfigure}

    \vspace{0.2cm}

    \begin{subfigure}[b]{0.238\textwidth}
        \centering
        \hspace{0.4cm} \includegraphics[width=3.25cm]{./Figures/Legends.eps}
        \vspace{0.03cm}
        \centering
        \includegraphics[width=\linewidth]{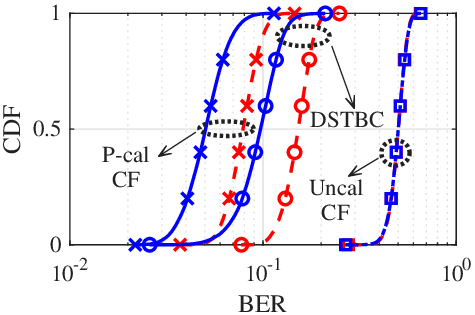}
        \caption{BER, $L_k = 2$ \acp{AP}}
        \label{Eq:CDF_c}
    \end{subfigure}
    \hspace{-0.31cm}
    \begin{subfigure}[b]{0.238\textwidth}
        \centering
        \hspace{0.4cm} \includegraphics[width=3.25cm]{./Figures/Legends.eps}
        \vspace{0.03cm}
        \centering
        \includegraphics[width=\linewidth]{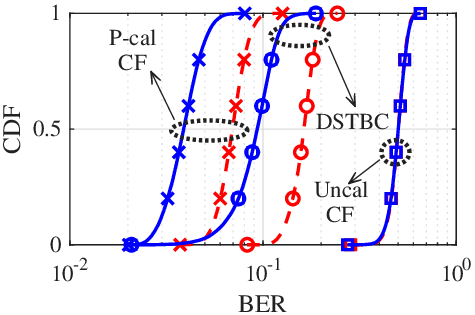}
        \caption{BER, $L_k = 4$ \acp{AP}}
        \label{Eq:CDF_d}
    \end{subfigure}

    \caption{\Ac{CDF} of the average \ac{SE} and \ac{BER} for each setup. Here: $L = 40$, $K = 20$, $N_{\mathrm{AP}} = 8$, $N_{\mathrm{UE}} = 2$, and $N_{\mathrm{s}} = 2$.}
    \label{Fig:CDF}
\end{figure}

Fig.~\ref{Fig:CDF} shows the \acp{CDF} of the average \ac{BER} and \ac{SE} for perfectly calibrated (P-cal) and uncalibrated (Uncal-CF) \ac{CF-mMIMO} systems, compared with the proposed approach (DSTBC) under both \ac{ZISI} and \ac{P-MMSE} precoding. Phase offsets from the \ac{UE} antennas degrade the \ac{BER} and \ac{SE} of \ac{CF-mMIMO} systems with coherent \ac{DL} transmission. 
In the uncalibrated case, \ac{P-MMSE} and \ac{ZISI} perform similarly because \ac{UE}-side phase offsets break channel reciprocity, preventing accurate interference mitigation and phase alignment. In the  DSTBC-based scheme, \ac{P-MMSE} still outperforms ZISI thanks to better interference suppression, which helps the proposed method counteract the unwanted cross-product effects in \eqref{eq:DifferentialDecodingTraceRealMin}. However, the \ac{SE} gain decreases for $L_k > 2$ due to the DSTBC pre-log factor, which depends on the code rate $R$ ($R = 3/4$ for $L_k = 4$) \cite{freitas2025PMCellFree}. This reveals an \ac{SE}–\ac{BER} trade-off: larger \ac{AP} clusters improve reliability (\ac{BER}) via diversity but incur an \ac{SE} penalty, requiring a careful choice of $L_k$ and the number of data streams.

\begin{figure}[t!]
    \centering
    \begin{subfigure}[b]{.23\textwidth}
        \centering
        \hspace{0.25cm} \includegraphics[width=3.25cm]{./Figures/Legends.eps}
        \includegraphics[width=\textwidth]{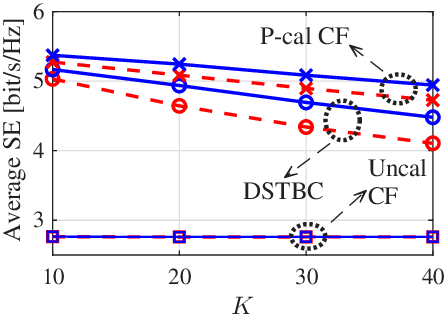} 
        \caption{$N_{\mathrm{UE}} = 2$}
        \label{Fig:CDF_SE_LP_MMSE_N_10_alamouti}
    \end{subfigure}
    \hspace{-0.2cm}
    \begin{subfigure}[b]{.232\textwidth}
        \centering
        \hspace{0.3cm} \includegraphics[width=3.25cm]{./Figures/Legends.eps}
        \includegraphics[width=\textwidth]{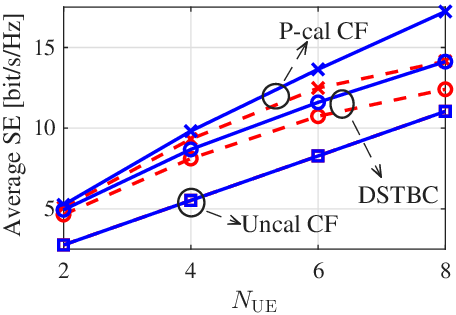} 
        \caption{$K = 20$}
        \label{fig:CDF_SE_P_MMSE_N_10_alamouti}
    \end{subfigure}

    \caption{Average \ac{SE} as a function of (a) the number of \acp{UE} $K$ in the network and (b) the number of antennas per \ac{UE}, $N_{\mathrm{UE}}$. Here, $L \!=\! 40$, $N_{\mathrm{AP}} \!=\! 8$, $L_k \!=\! 2$, and $N_s = N_{\mathrm{UE}}$.}
    \label{Fig:averageSE}
\end{figure}

Fig.\,\ref{Fig:averageSE}, shows the average network \ac{SE} versus the number of \acp{UE} $K$ and antennas per \ac{UE} $N_{\mathrm{UE}}$. The proposed approach restores the \ac{SE} of \ac{UC} \ac{CF-mMIMO} systems as $K$ and $N_{\mathrm{UE}}$ increase, whereas \ac{ZISI} degrades more with larger $K$ due to weaker interference suppression than \ac{P-MMSE}. The \ac{SE} does not grow linearly with $N_{\mathrm{UE}}$ because extra data streams increase inter-stream interference and pilot contamination worsens with $N_{\mathrm{UE}}$, limiting \ac{SE} gains. Finally, although not shown due to space constraints, the proposed approach also improves the SE across different numbers of UL pilots and various levels of amplitude and phase variations in the UE RF chains, and is particularly effective for larger phase variations.

\section{Conclusions}
This letter introduced a \ac{DSTBC}-based strategy to mitigate the impact of UE antenna miscalibration in  \ac{CF-mMIMO} systems. 
Results showed that uncalibrated arrays at the \ac{UE} degrade \ac{BER} and \ac{SE} in conventional setups, while the proposed scheme approaches the ideal calibrated case. 
Future work includes investigating reliable OTA calibration schemes for compensating UE-side calibration errors.

\section*{Appendix A: the \ac{P-MMSE} Precoder}

The \ac{P-MMSE} precoding suppresses interference from UEs that mostly affect UE~$k$ DL transmission, defined as $\mathcal{P}_{k} = \{i : \mathbf{a}_{k} \mathbf{a}_{i}^{\mathrm{H}} \neq \mathbf{0}_{L \times L} \}$, where $\mathbf{a}_k = [\mathrm{a}_{k,1}, \ldots, \mathrm{a}_{k,L}]^{\mathrm{T}} \in \mathbb{C}^{L \times 1}$. It is obtained from the MMSE formulation in \cite{kama2025multiantennausers}, but applied only to \acp{UE} in $\mathcal{P}_k$. Denoting the estimated channel matrix of \ac{UE}~$i$ as $\mathbf{\widehat{G}}_{i}^{\mathrm{UL}} = [ (\mathrm{a}_{i,1} \mathbf{\widehat{G}}_{i, 1}^{\mathrm{UL}})^{\mathrm{T}}, \ldots, (\mathrm{a}_{i,L} \mathbf{\widehat{G}}_{i, L}^{\mathrm{UL}})^{\mathrm{T}} ]^{\mathrm{T}} \in \mathbb{C}^{LN_{\mathrm{AP}}\times N_{\mathrm{UE}}}$, the \ac{P-MMSE} precoder is $\mathbf{W}_{k}^{\mathrm{P-MMSE}} = \sqrt{\rho_k} \eta_{k} \! \left[ \sum_{i \in \mathcal{P}_k} \eta_{i} \mathbf{\overline{W}}_{i} +  \sigma_{\mathrm{UL}}^{2}\mathbf{I}_{LN_{\mathrm{AP}}} \right]^{-1} \!\mathbf{\widehat{G}}_{k}^{\mathrm{UL}}$,
where $\mathbf{\overline{W}}_{i} = \mathbf{\widehat{G}}_{i}^{\mathrm{UL}} (\mathbf{\widehat{G}}_{i}^{\mathrm{UL}} )^{\mathrm{H}} + \mathbf{Z}_{\mathbf{\bar{G}}_{i}^{\mathrm{UL}}}$, with $\mathbf{Z}_{\mathbf{\bar{G}}_{i}^{\mathrm{UL}}} = \mathbb{E}\{ \mathbf{\bar{G}}_{i}^{\mathrm{UL}} (\mathbf{\bar{G}}_{i}^{\mathrm{UL}})^{\mathrm{H}} \}$ being the covariance of the estimation error $\mathbf{\bar{G}}_{i}^{\mathrm{UL}} = \mathbf{\widetilde{G}}_{i}^{\mathrm{UL}} - \mathbf{\widehat{G}}_{i}^{\mathrm{UL}}$. Here, $\rho_k$ is the total power allocated to \ac{UE} $k$, $\eta_k$ is its per-antenna transmit power, and $\sigma_{\mathrm{UL}}^{2}$ is the \ac{UL} noise variance.

\section*{Appendix B: UE Calibration Error Mitigation}
We now give a simplified analytical example under specific assumptions to show how DSTBC schemes mitigate UE-side antenna calibration errors. Assuming perfect channel estimates at the APs, a ZISI precoder, and perfectly calibrated AP antenna arrays (i.e., $\mathbf{\Phi}_{rx,l}^{\mathrm{AP}} = \mathbf{\Phi}_{tx,l}^{\mathrm{AP}} = \mathbf{I}_{N_\mathrm{AP}}$), and following the same procedure as for \eqref{eq:usefuldata_uncalibrated}, Eq.~\eqref{eq:received_block} can be rewritten as
\begin{equation}
    \mathbf{Y}^t_k =\sum\nolimits_{l \in \mathcal{L}_k} \sqrt{\rho_{k,l}} \widetilde{\mathbf{\Phi}}_{rx,k}^{\mathrm{UE}} \big[\widetilde{\mathbf{\Phi}}_{tx,k}^{\mathrm{UE}} \big]^\mathrm{-H}\, \mathbf{M}_k\, \mathbf{B}^t_{k,l} \;+\; \mathbf{N}^t_k \,.
\end{equation}
W.l.o.g., assume $N_{\mathrm{UE}} = 4$, $N_s = 2$, and $L_k = 2$. By denoting the indexes of the serving APs as $l$ and $l'$, the useful term $\mathbf{U}^{t}_k$ of $\mathbf{Y}^t_k$ becomes
\begin{equation}
\mathbf{U}^{t}_k = \begin{bmatrix}
\phi_{r,k}^1 (\phi_{t,k}^1)^{-\mathrm{H}} \mathbf{c}^t_{k,l,1}\\
\phi_{r,k}^2 (\phi_{t,k}^2)^{-\mathrm{H}} \mathbf{c}^t_{k,l,1}\\
\phi_{r,k}^3 (\phi_{t,k}^3)^{-\mathrm{H}} \mathbf{c}^t_{k,l,2}\\
\phi_{r,k}^4 (\phi_{t,k}^4)^{-\mathrm{H}} \mathbf{c}^t_{k,l,2}
\end{bmatrix} +
\begin{bmatrix}
\phi_{r,k}^1 (\phi_{t,k}^1)^{-\mathrm{H}} \mathbf{c}^t_{k,l',1}\\
\phi_{r,k}^2 (\phi_{t,k}^2)^{-\mathrm{H}} \mathbf{c}^t_{k,l',1}\\
\phi_{r,k}^3 (\phi_{t,k}^3)^{-\mathrm{H}} \mathbf{c}^t_{k,l',2}\\
\phi_{r,k}^4 (\phi_{t,k}^4)^{-\mathrm{H}} \mathbf{c}^t_{k,l',2}
\end{bmatrix}.
\end{equation}
To detect stream $j$ at UE $k$,  $\mathbf{\widetilde{Y}}^t_{k,j}$ is computed as in \eqref{eq:stream_extraction}. 
We focus now on the useful term of $\mathbf{\widetilde{Y}}^t_{k,j}$ and neglect the interference-plus-noise terms $\mathbf{N}^t_k \in \mathbb{C}^{N_{\mathrm{UE}} \times L_k}$ for analytical tractability. Thus, for the first stream ($j = 1$), we can write $\mathbf{\widetilde{U}}^t_{k,1} \in \mathbb{C}^{N_b \times L_k}$ from $\mathbf{\widetilde{Y}}^t_{k,j}$ as
\begin{equation}
\begin{bmatrix}
\phi_{r,k}^1 (\phi_{t,k}^1)^{-\mathrm H} & 0 \\
0 & \phi_{r,k}^2 (\phi_{t,k}^2)^{-\mathrm H}
\end{bmatrix}
\begin{bmatrix}
\mathbf{c}^t_{k,l,1} + \mathbf{c}^t_{k,l',1} \\
\mathbf{c}^t_{k,l,1} + \mathbf{c}^t_{k,l',1}
\end{bmatrix}
\end{equation}
and define $\mathbf{\widetilde{U}}^{t-1}_{k,1}$ in a similar manner. By performing $( \mathbf{\widetilde{U}}^t_{k,1} )^{\mathrm{H}} \mathbf{\widetilde{U}}^{t-1}_{k,1}$, we obtain
\begin{equation}
\begin{bmatrix}
\mathbf{c}^{t}_{k,l,1} + \mathbf{c}^{t}_{k,l',1} \\
\mathbf{c}^{t}_{k,l,1} + \mathbf{c}^{t}_{k,l',1}
\end{bmatrix}^{\mathrm{H}}
\mathbf{\widetilde{A}}_{r,k}^{\mathrm{UE}}
\begin{bmatrix}
\mathbf{c}^{t-1}_{k,l,1} + \mathbf{c}^{t-1}_{k,l',1} \\
\mathbf{c}^{t-1}_{k,l,1} + \mathbf{c}^{t-1}_{k,l',1}
\end{bmatrix}.
\end{equation}
where $\mathbf{\widetilde{A}}_{r,k}^{\mathrm{UE}} = \mathrm{diag} \big( |\phi_{r,k}^1 (\phi_{t,k}^1)^{-\mathrm H}|^{2} \;, \; |\phi_{r,k}^2 (\phi_{t,k}^2)^{-\mathrm H}|^{2} \big)$. Since $\mathbf{c}^{t}_{k,l,j}$ can also be regarded as $\mathbf{c}^{t}_{k,l,j} = \mathbf{c}^{t-1}_{k,l,j}\mathbf{X}^{t}_{k,j}$ in \eqref{eq:row_mapping}, the product $( \mathbf{\widetilde{U}}^t_{k,1} )^{\mathrm{H}} \mathbf{\widetilde{U}}^{t-1}_{k,1}$ becomes 
\begin{equation}
\mathbf{X}^{t}_{k,1} [\mathbf{c}^{t-1}_{k,l,1} + \mathbf{c}^{t-1}_{k,l',1}]^{\mathrm{H}} [\mathbf{c}^{t-1}_{k,l,1} + \mathbf{c}^{t-1}_{k,l',1}] \| \mathbf{\widetilde{A}}_{r,k}^{\mathrm{UE}} \|_{F}^{2},
\end{equation}
where it can be observed that the effects of UE-side antenna calibration errors are effectively mitigated since $\| \mathbf{\widetilde{A}}_{r,k}^{\mathrm{UE}} \|_{F}^{2} = (|\phi_{r,k}^1 (\phi_{t,k}^1)^{-\mathrm H}|^{2} + |\phi_{r,k}^2 (\phi_{t,k}^2)^{-\mathrm H}|^{2})$ is constant during the time interval of two consecutive codewords. The same reasoning applies to the second stream ($j = 2$).

\bibliographystyle{IEEEtran}
\bibliography{IEEEabrv,bib_file}

\vfill

\end{document}

%% file: acronym.tex
\begin{acronym}
    \acro{5G}{fifth-generation}
	\acro{AP}{access point}
	\acro{AWGN}{additive white Gaussian noise}
	\acro{ASD}{Angular standard deviation}
	\acro{ABC}{Artificial Bee Colony}
	\acro{AoA}{angle of arrival}
    \acro{ADC}{analog-to-digital converter}
	\acro{B5G}{beyond 5G}
	\acro{BA}{Bat Algorithm}
	\acro{BS}{base station}
	\acro{BER}{bit-error rate}
	\acro{BPSK}{binary phase shift keying}
	\acro{CS}{Cuckoo Search}
	\acro{CAPEX}{capital expenditure}
	\acro{CF}{cell-free}
    \acro{CF-mMIMO}{cell-free massive multiple-input multiple-output}
	\acro{CPU}{central processing unit}
	\acro{CC}{computational complexity}
    \acro{CM}{complex multiplications}
	\acro{CTMC}{continuous-time Markov chain}
	\acro{CSI}{channel state information}
	\acro{CSIR}{CSI at the receiver}
	\acro{CSIT}{CSI at the transmitter}
	\acro{CDF}{cumulative distribution function}
	\acro{CHD}{channel hardening}
	\acro{DL}{downlink}
	\acro{DCC}{dynamic cooperation clustering}
	\acro{D-mMIMO}{distributed massive multiple-input multiple-output}
    \acro{DFT}{discrete Fourier transform}
    \acro{DSP}{digital signal processor}
    \acro{DPSK}{differential phase-shift keying}
    \acro{DSTBC}{differential space-time block coding }
	\acro{EE}{energy efficiency}
	\acro{FS}{fiber switch}
	\acro{FD}{full duplication}
        \acro{FFT}{fast Fourier transform}
	\acro{FVP}{favorable propagation}
	\acro{FPA}{flower pollination algorithm}
	\acro{FA}{Firefly Algorithm}
	\acro{GA}{Genetic Algorithm}
        \acro{GOPS}{giga operations per second}
	\acro{GWO}{Grey Wolf Algorithm}
 \acro{GPP}{general purpose processor}
	\acro{HCPP}{hard core point process}
	\acro{i.i.d}{independent and identically distributed}
	\acro{InH-open}{Indoor Hotspot Open Office}
	\acro{IoT}{Internet of Things}
	\acro{IC}{inter-coordinated}
	\acro{LOS}{line-of-sight}
	\acro{LP-MMSE}{local partial MMSE}
	\acro{LSFD}{large‐scale fading decoding }
	\acro{LSFB}{Largest-large‐scale fading}
    \acro{LMMSE}{linear minimum mean square error} 
    \acro{ML}{maximum likelihood}
	\acro{MD}{matched-decision}	
	\acro{MCMC}{markov chain monte carlo}
	\acro{MIMO}{multiple-input multiple-output}
	\acro{MF}{matched filter}
	\acro{m-MIMO}{massive-multiple-input multiple-output}
	\acro{MTBF}{mean time between failures}
	\acro{MR}{maximum ratio}
	\acro{MOFPA}{Multiobjective Flowers Pollination Algorithm}
	\acro{MMSE}{minimum mean square error}
	\acro{NLOS}{non-line-of-sight}
    \acro{NCC}{network-centric}
	\acro{NMSE}{normalized mean square error}
	\acro{NOMA}{non-orthogonal multiple access}
	\acro{NP}{no protection}
	\acro{NS}{non-scalable}
	\acro{NF}{noise figure}
    \acro{NR}{new radio}
	\acro{OMA}{orthogonal multiple access}
	\acro{OFDM}{orthogonal frequency-division multiplexing}
	\acro{OPEX}{operational expenditure}
    \acro{OSTBC}{orthogonal STBC}
    \acro{OTA}{over-the-air}
	\acro{PRBs}{physical resource blocks}
	\acro{PD}{partial duplication}
	\acro{P-MMSE}{partial MMSE}
	\acro{P-RZF}{partial regularized zero-forcing}
	\acro{PDF}{probability density function}
	\acro{PSO}{particle swarm optimization}
    \acro{PA}{proposed approach}
    \acro{PEP}{pairwise error probability}
    \acro{PSK}{phase-shift keying}
	\acro{QAM}{quadrature amplitude modulation}
	\acro{QoS}{quality-of-service}
	\acro{RF}{radio frequency}
	\acro{RSMA}{rate-splitting multiple access}
	\acro{RMSE}{root-mean-square deviation}
	\acro{RV}{random variable}
	\acro{RS}{radio Stripes}
    \acro{RU}{radio unit}
	\acro{SB}{serial buse}
	\acro{SDMA}{space-division multiple acess}
	\acro{SE}{spectral efficiency}
	\acro{SFP} {small form-factor pluggable}
	\acro{SHR}{self-healing radio}
	\acro{SIC}{successive interference cancellation}
	\acro{SCF}{scalable cell-free}
	\acro{SINR}{signal-to-interference-plus-noise ratio}
	\acro{SNR}{signal-to-noise ratio}
    \acro{STBC}{space-time block coding}
	\acro{TCO}{total cost of ownership}
	\acro{TDD}{time-division duplex}
	\acro{TR}{Technical Report}
  	\acro{TRP}{transmission and reception point}
	\acro{UatF}{use-and-then-forget}
	\acro{UAV}{unmanned aerial vehicle}
	\acro{UE}{user equipment}
	\acro{UL}{uplink}
	\acro{UMi}{Urban Micro}
	\acro{UC}{user-centric}
    \acro{UCC}{User-centric clustering}
    \acro{ULA}{uniform linear array}
    \acro{UPA}{uniform planar array}
    \acro{ZISI}{zero inter-stream interference}
\end{acronym}